\documentclass[12pt,a4paper]{llncs}

\usepackage[utf8]{inputenc} 
\usepackage{cite}
\usepackage{amsmath,amssymb,amsfonts}
\usepackage{algorithmic}
\usepackage{graphicx}
\usepackage{textcomp}
\usepackage{xcolor}
\usepackage{stfloats}
\usepackage{float}
\usepackage{subfig}
\usepackage{adjustbox}
\usepackage[labeled]{multibib}

    \newcommand{\Tuple}[1]{\ensuremath{\langle #1\rangle}}
     \newcommand{\bTuple}[1]{\ensuremath{\langle {\bf #1}\rangle}}
    \newcommand{\nbTuple}[2]{\ensuremath{\langle {#1},{\bf #2}\rangle}}
    
\begin{document}

\mainmatter  

\title{Multiple Sclerosis disease: a computational approach for investigating its drug interactions.}

\titlerunning{Multiple Sclerosis disease: a computational approach for investigating its drug interactions.}

%
%

\author{Simone Pernice$^{(1)}$, Marco Beccuti$^{(1)}$, Greta Romano$^{(1)}$, Marzio Pennisi$^{(3)}$, Alessandro Maglione$^{(2)}$, Santina Cutrupi$^{(2)}$, Francesco Pappalardo$^{(4)}$, Lorenzo Capra$^{(5)}$, Giuliana Franceschinis$^{(6)}$, Massimiliano De Pierro$^{(1)}$, Gianfranco Balbo$^{(1)}$, Francesca Cordero$^{(1)}$, Raffaele Calogero$^{(7)}$  }

\institute{$\ $\\(1) Dept. of Computer Science, University of Turin, Turin, Italy\\
(2) Dept. of Clinical and Biological Sciences, University of Turin, Orbassano, Italy\\
(3) Dept. of Mathematics and Computer Science, University of Catania, Catania, Italy\\
(4) Dept. of Drug Sciences, University of Catania, Catania, Italy\\
(5) Dept. of Computer Science, University of Milan, Milan, Italy \\
(6) Computer Science Inst., DiSIT, University of Eastern Piedmont, Alessandria, Italy \\
(7) Dept. of Molecular Biotechnology and Health Sciences, University of Turin, Turin, Italy\\
}

\maketitle

\begin{abstract}

Multiple Sclerosis (MS)  is a chronic and potentially highly disabling disease that can cause permanent damage and deterioration of the central nervous system. In Europe it is the leading cause of non-traumatic disabilities in young adults,  since  more than 700,000 EU people suffer from MS.
Although recent studies on MS pathophysiology have been provided,  MS remains a challenging disease.
In this context, thanks to recent advances in 
software and hardware technologies, computational models and computer simulations are becoming  appealing  research tools to support scientists in the study of such disease. Thus, motivated by this consideration  we  propose in this paper a new model to study the evolution of MS in silico, and the effects of the administration of  Daclizumab  drug, taking into account also  spatiality and temporality of the involved phenomena. Moreover, we  show how the  intrinsic symmetries of the system can be exploited to drastically reduce the complexity of its analysis.



\end{abstract}

\section{\bf Introduction}
Multiple Sclerosis (MS) is a long-term and autoimmune disease of the Central Nervous System (CNS). During the progression of the disease, cells of immune system attack the principal components of the CNS, the neurons, removing the enveloping myelin and preventing the efficient transmission of the nervous signals.
Relapsing-Remitting MS (RRMS) is the predominant type of MS since it is diagnosed in about $85\% -90\%$ of MS cases \cite{Sospedra2005}. 
In RRMS, the disease alternates two phases: 
(1) relapse phase is characterized by a disease worsening due to the active inflammation damaging the neurons; (2) in the remission phase there is a complete or partial lack of the symptoms \cite{Dutta2011}.
Recently, many treatments were proposed and studied to contrast the RRMS progression. Among these drugs, daclizumab \cite{Bielekova2019} (commercial name Zynbrita), an antibody tailored against the Interleukin-2 receptor (IL2R) of T cells, exhibited promising results. Unfortunately, its efficacy was accompanied by an increased frequency of serious adverse events as  
infections, encephalitis, and liver damages.
For these reasons daclizumab has been withdrawn from the market worldwide.

In ~\cite{PerniceBMC2019} we proposed a model to investigate the effect of the daclizumab administration in RRMS. It involves the following seven main actors of MS: Epstain-Barr virus (EBV), Effector T lymphocytes cells (Teff), Regulatory T lymphocytes cells (Treg), Natural Killer cells (NK), Oligodentrocytes  cells (ODC), Interleukin-2 (IL2) and daclizumab (DAC).
In details, the EBV was considered  since several studies \cite{Virtanen2012} commonly agree with the hypothesis that viruses may play a role in RRMS pathogenesis acting as environmental triggers, and in particular the presence of this virus represents a well established risk factor in MS \cite{Guan2019}.
Effector T cells (Teff) are instead immune cells with a protective role against pathogens in healthy people. However, in RRMS a hypothesis is that the EBV first infection could bring to the activation of autologous Teff lymphocytes against myelin, due to a structure similarity between one viral protein and myelin protein (molecular mimicry). Regulatory T cells (Treg)  are immune cells acting as balancing of the immune response since they contribute to suppress and modulate the Teff cells activity when no longer needed, or when there is a high risk of inflammation that can cause injuries to the tissues of the host.
Another important actor are the natural killer (NK) cells, a family of immune cells that acts as host-rejection of infected cells.
Oligodendrocytes (ODC) are instead cells supporting the neurons since they produce and are able to partially restore the myelin around the neurons if a not excessive damage occurs.
IL2 is an immunomodulatory cytokine released by Teff in order to self-stimulate to duplicate and to propagate their immune actions.
Finally, we included in the model the drug daclizumab, a humanized monoclonal antibody used in MS as drug against  the Interleukin-2 receptor (IL2R) that is able to break  the autoimmune reaction by suppressing the immune cells proliferation \cite{Bielekova2019}.

Thus, to help scientists in improving their knowledge of these phenomena, in this work we extend the RRMS models presented in ~\cite{PerniceBMC2019}  considering the cells movement into a three-dimensional grid.
In details, in this paper we show how the use of a graphical formalism, i.e.  the Extended Stochastic Symmetric Net (ESSN) formalism ~\cite{PerniceBMC2019,FondamentaeInfo},  allows us to easily deal with  this  complex three-dimensional model  whose direct definition in terms of ODE system  becomes clearly unfeasible even  for a small  three-dimensional grid.\\
Indeed, for instance considering  three-dimensional grid with dimension $3\times3\times3$  the ESSN model is  a bipartite  graph with only  38 nodes (i.e. 13 places and 25 transitions) and  approximately 90 arcs; while  its  underlying deterministic process is an ODE system with 433 equations.\\
Moreover, the high level of parametrization and flexibility provided in the model through this graphical formalism enables us to study different grid dimensions without changing the structure of the model.
Similarly, in the analysis phase the ESSN model provides a powerful methodology  that  automatically exploits  the system symmetries to reduce the complexity (in terms of number of equations) of  the underlying deterministic process.
Indeed in~\cite{Beccuti201936} we proposed an algorithm that directly derives a \emph{compact} ODE system from a ESSN model in a symbolic way, through algebraic manipulation of ESSN annotations.

\section{\bf Scientific background}
In this section we introduce the Petri Nets (PNs) formalism used to describe our model. PNs and their extensions  are   effective formalisms to model biological systems thanks to their capability of representing in a simple and clear  manner the system features and to provide efficient techniques to derive system qualitative and quantitative properties.
In details, PNs are bipartite directed graphs with two types of nodes called \emph{places} and \emph{transitions}. Places, graphically represented as circles, correspond to the state variables of the system, while transitions, graphically represented as boxes, correspond to the events that can induce a state change. The arcs connecting places to transitions and vice versa express the relations between states and event occurrences. Places can contain tokens, drawn as black dots. The state of a PN, namely \emph{marking}, is defined by the number of tokens in each place.
The system evolution  is  provided by the firing of an enabled transition, where a transition is enabled if and only if each input place contains a number of tokens greater than or equal to a given threshold defined by the cardinality of the corresponding input arc.
The firing of an enabled transition removes a fixed number of tokens from its input places and adds a fixed number of tokens into its output places (according to the cardinality of its input/output arcs).

In this work we focus on   Stochastic Symmetric Nets (SSNs) a high level formalism that extends PNs  with \emph{colors} and \emph{stochastic firing delays}~\cite{CDFH93}. Colors provide a more compact, readable and parametric representation of the system thanks to the possibility of having distinguished tokens.

More specifically, the \emph{color domain} associated with a place specifies  
the color of the  tokens contained in
this place, whereas the color domain of a transition defines the different ways of firing it (i.e. the possible \emph{transition instances}).
In order to specify these firings, a color function is attached to each arc which,
given a color of the transition connected to the arc, determines the number of colored tokens that will be added to or removed from the corresponding place.
A color domain is defined as Cartesian product of \emph{color classes} which may be viewed as primitive domains. A color class can be partitioned into \emph{static subclasses}. The colors of a class have the same nature (e.g. T cells), whereas the colors inside a static subclass have the same potential behavior (e.g. Teff).
Stochastic firing delays, sampled from a negative exponential distribution,  allow  to automatically derive the underlying Continuous Time Markov Chain (CTMC) that can be studied to quantitatively evaluate the system behaviour. In the literature, different techniques are proposed to solve the underlying CTMC; in particular, in case of very complex models, the so-called deterministic approach~\cite{Ku70} can be efficiently exploited. According to this, in~\cite{Beccuti15} we proposed how to derive a deterministic process, described through a system of Ordinary Differential Equations (ODEs), which well approximates the stochastic behavior of an SSN model assuming all reactions follow the Mass Action (MA) law. 
In the same paper we also described an efficient translation method based on the SSN formalism, which is able to reduce the size (in terms of equations number) of the underlying  ODE system through the automatic exploitation of system symmetries.
Practically, the complete set of ODEs,  which can be
derived from  an SSN model is  partitioned  into equivalence classes of ODEs which have same solution so that a representative equation,  called \emph{symbolic equation},  can be pointed out for each equivalence class.
Then, a reduced ODE system may be derived including only these symbolic equations whose solution mimics the behavior of the original model. 
Recently this result  was further improved in~\cite{Beccuti201936} where  a new algorithm is discussed  to generate  the symbolic equation for each equivalence class of ODEs without  deriving the complete ODE system. This is achieved thanks to  a recent  extension of a symbolic calculus for the computation of SSN structural properties~\cite{CapraPF15}.\\
Furthermore, in~\cite{PerniceBMC2019} we introduced 
the Extended SSNs (ESSNs) to deal with more complex biological laws splitting the set $T$ of all the transitions into two subsets: $T_{ma}$ and $T_g$. 
Thus, the former subset contains transitions (that are called {\em standard}) whose rates are  specified as MA laws. The latter includes instead all the transitions (that are called {\em  general}) whose random firing times have rates that are defined as general real functions. 
In our definition, we assumed that the general function associated with a transition $t \in T_g$ is a real function which depends only on time and on the input places of $t$. So, if $x_{p,c}(\nu)$ represents the average number of tokens  of color $c$ in the place $p$ at time $\nu$, then the rate at which the instance $\langle t, c, c'\rangle$, $t \in T_g$  will move tokens with color $c$ in  place  $x_{p,c'}(\nu)$ is given by $f_{\langle t,c , c'\rangle}(\hat x(\nu), \nu)$, where $\hat x(\nu)$ is the vector  characterized by the average number of tokens of the input places of transition $t$.

\section{\bf Materials and Methods}
\begin{figure}[tbp]
    \centering
    \includegraphics[width=1\textwidth]{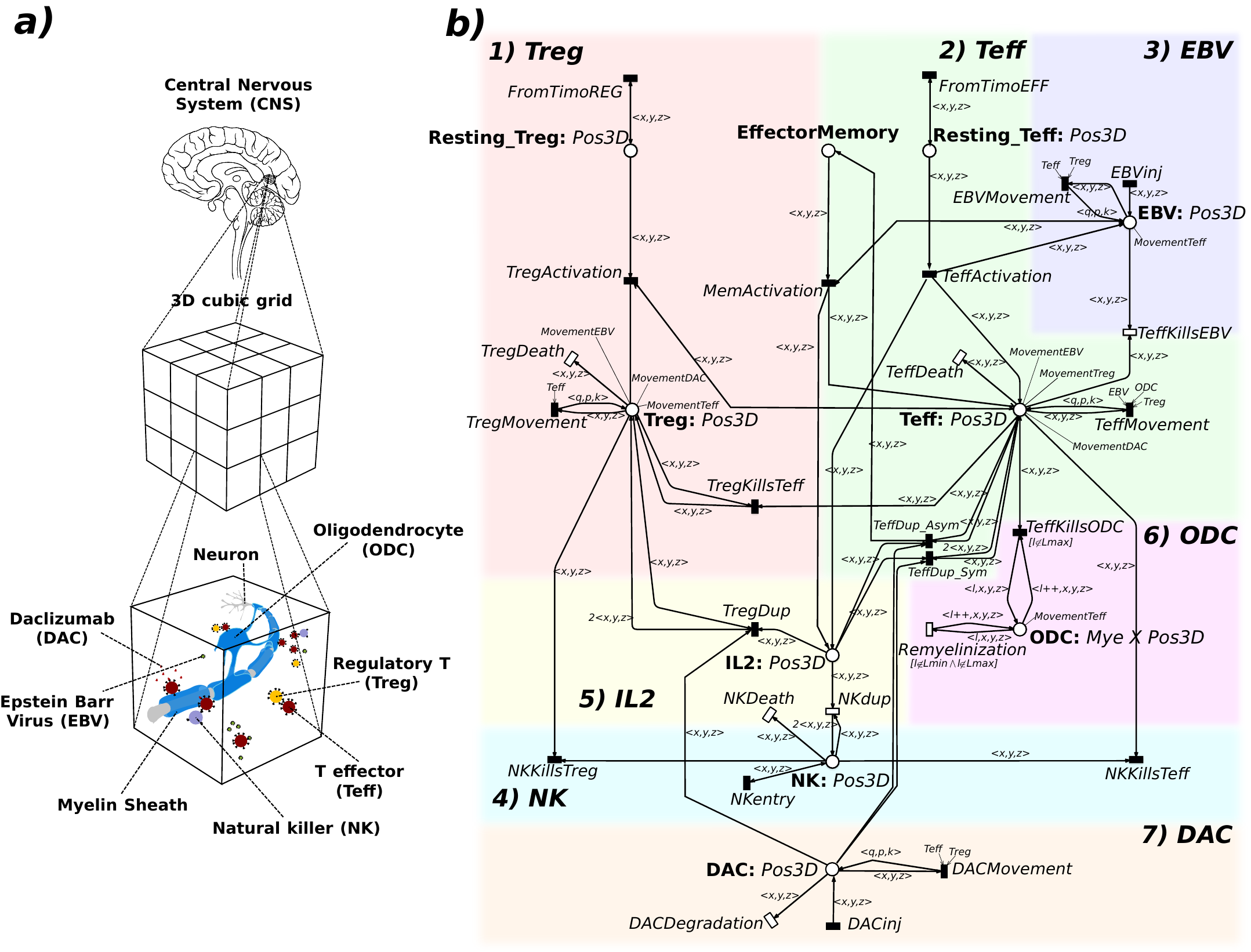}   
    \caption{ a) Representation of the three-dimensional model. Each cube contains the 7 elements characterizing the MS disease. Elements position is identified by three coordinates in the 3D cubic grid, representing a piece of CNS b) The ESSN model.}
    \label{fig:RRMSModel}
\end{figure}
In this section, we report our extension of the Relapsing-Remitting Multiple Sclerosis (RRMS) model presented in \cite{PerniceBMC2019} considering the cell movements in a cubic grid. The model, shown in Fig.~\ref{fig:RRMSModel}b), consists of  13 places and 25 transitions. For the sake of clarity, the white transitions are standard ones, while the black  ones are  general transitions. All the general functions, the constants and numerical values associated with the transitions, except for those regarding the movement transitions, are reported in the \cite{PerniceBMC2019}'s Supplementary file.
This model is organized in seven modules corresponding to the biological entities characterizing RRMS. Briefly, the EBV module simulates the virus injections in the system, while the Treg and Teff modules  encode the activation of the T cells, the annihilation of the virus by the Teff action, the control mechanism of the Treg over the Teff. The NK module describes the killing of self-reactive Teff and Treg cells respectively, due to NK cells. The IL2 module is focused on the IL2 role. IL2 is consumed by the Treg, Teff and NK functions and it is produced by the Teff activation.  The ODC module describes instead the ODC behaviour, characterized particularly by the damage caused by Teff cells on ODC cells. Indeed, when the myelin level reaches the lowest value, an irreversible damage occurs and a remyelination of the neurons is no more possible.

Finally, the DAC module encodes the drug administration and its pharmacokinetics inhibition of the expansion of Treg and Teff.

The model is characterized by four color classes: \emph{PosX}, \emph{PosY}, and \emph{PosZ} representing the coordinates of the position of a molecule in a 3D cubic grid;  \emph{Mye} encoding the myelination levels of ODC. \emph{Mye} is divided into five static subclasses ranging from \emph{Lmin} (no myelination) to \emph{Lmax} (full myelination).
Then, all the places except the ODC and EffectorMemory are characterized by the color domain defined as $Pos3D = PosX \times PosY  \times PosZ$, i.e. the three-dimensional Cartesian product of the three coordinates color classes. Instead, the ODC place is characterized by the three coordinates plus the myelination levels, so that its color domain is $Pos3D \times Mye$. Finally,  the EffectorMemory place has neutral color domain.

Moreover, we assume that the EBV, Teff, Treg and DAC cells are able to move in all the  cubic cells of the grid. Practically,  the EBVs move uniformly in all the cells, the Teff cells  move with higher probability towards a location in which there is higher concentration of EBV, and Treg and DAC  cells move with higher probability towards a location in which there is higher concentration of Teff cells. Hereafter, the notation of the color combinations  $\Tuple{p_x,p_y,p_z}$ and $\Tuple{q_x,q_y,q_z}$, representing the location coordinates,  is simplified to $\bTuple{p}$ and $\bTuple{q}$, respectively. In particular, we define $x_{CellType_{\bTuple{p}}}$ as the number of $CellType$ in the location  $\bTuple{p}$ at a specific time point. Hence, the movement functions are defined as follows:\\


\textbf{\emph{TeffMovement}} simulates the Teff movements from the coordinates represented by the color combination $\bTuple{p}$ to $\bTuple{q}$. Its velocity is inversely proportional to the number of EBV in  $\bTuple{p}$ and depends on the number of EBV in  $\bTuple{q}$ such that a greater number of EBV cells leads to  a higher probability to move into that location.
\begin{align*}
    f_{\nbTuple{TeffMovement}{p,q}} (\hat{x}(\nu),\nu)=& r_{moves} * ( exp(- \dfrac{x_{EBV_{\bTuple{p}}} }{C_{EBV}} ) ) * p^{Teff}_{\bTuple{q}} * x_{Teff_{\bTuple{p}}} 
\end{align*}
where $p^{Teff}_{\bTuple{q}}= \frac{x_{EBV_{\bTuple{q}}}}{EBV_{tot}}$ represents the probability to move in the cell with coordinates $\bTuple{q}$ and $EBV_{tot}$ is the total number of EBV in the grid at time $\nu$. Moreover, in our experiment we fixed $r_{moves}=0.1$ and $C_{EBV}=1000$. \\
\\
\textbf{\emph{TregMovement}} represents the Treg movements from the coordinates represented by the color combination $\bTuple{p}$ to the coordinates $\bTuple{q}$. This is inversely proportional to the number of Teff in  $\bTuple{p}$ and depends on the number of Teff in  $\bTuple{q}$, such that a greater number of Teffs leads to  a higher probability to reach that location.

\begin{align*}
    f_{\nbTuple{TregMovement}{p,q}} (\hat{x}(\nu),\nu)=& r_{moves} * ( exp(- \dfrac{x_{Teff_{\bTuple{p}}}}{C_{\it Teff} }) ) * p^{Treg}_{\bTuple{q}} * x_{Treg_{\bTuple{p}}} 
\end{align*}
where $p^{Treg}_{\bTuple{q}}= \frac{x_{Teff_{\bTuple{q}}}}{Teff_{tot}}$ represents the probability to move in the cell with coordinates $\bTuple{q}$ and ${\it Teff}_{tot}$ is the total number of Teff in the grid at time $\nu$. Moreover, in our experiment we fixed $r_{moves}=0.1$ and $C_{\it Teff}=800$.\\
\\		
\textbf{\emph{EBVMovement}} simulates the EBV movements from the coordinates represented by the color combination $\bTuple{p}$ to the coordinates $\bTuple{q}$. In this case we assume that the probability to move is equally distributed among  the grid cells.

\begin{equation*}
    f_{\nbTuple{EBVMovement}{p,q}} (\hat{x}(\nu),\nu)= r_{moves} * p^{EBV}_{\bTuple{q}} * x_{EBV_{\bTuple{p}}} 
\end{equation*}

where $p^{EBV}_{\bTuple{q}}= 1 / N_{grid} $ represents the probability to move in the cell with coordinates $\bTuple{q}$ with $N_{grid} $ being  the number of cells in the grid. Moreover, in our experiment we fixed $r_{moves}=0.1$.
\\		
\textbf{\emph{DACMovement}} simulates the DAC movements from the coordinates represented by the color combination $\bTuple{p}$ to the coordinates $\bTuple{q}$. This is inversely proportional to the number of T-cells (Treg+Teff) in  $\bTuple{p}$ and directly proportional to the number of T-cells in  $\bTuple{q}$. 

\begin{align*}
    f_{\nbTuple{DACMovement}{p,q}} (\hat{x}(\nu),\nu)=& r_{moves} * ( exp(-\dfrac{x_{Tcells_{\bTuple{p}}}}{C_{Tcell}}) )* p^{DAC}_{\bTuple{q}} * x_{DAC_{\bTuple{p}}} 
\end{align*}

where $p^{DAC}_{\bTuple{q}}= \dfrac{x_{Tcells_{\bTuple{q}}}}{Tcells_{tot}}$ represents the probability to move in the cell with coordinates $\bTuple{q}$. Moreover, in our experiment we fixed $r_{moves}=0.1$ and $C_{\it Tcell}=1000$.


All the R files, general transitions definition, data generated and analyzed during this study, and the GreatSPN file of the net are freely available at \url{https://github.com/qBioTurin/ESSNandRRMS/} \url{tree/master/DeterministicModel/Multidimensional}.\\ The system of SODE has been generated with the SNespression tool (\url{http://di.unito.it/$\sim$depierro/SNexpression}) and integrated with the definition of the functions for the {\em general} transitions and the initial marking.

\section{\bf Results}

In this work we studied the RRMS considering a tissue portion explicitly  modeled through a cubic  grid composed by 27 cubic cells (Fig.~\ref{fig:RRMSModel}a) ).
To achieve  this,  we defined  the color classes $\it{PosX}=\{x_1,x_2,x_3\}$,   $\it{PosY}=\{y_1,y_2,y_3\}$ and $\it{PosZ}=\{z_1,z_2,z_3\}$. For all the simulations, we assumed 500 ODC with level $L_{max}$ of neuronal myelinization, 1687 resting Teff cells, 63 resting Treg cells, 375 NK cells and 1000 IL2 molecules, and zero cells in the other places (for more details see \cite{PerniceBMC2019}).\\
We exploited the GreatSPN tool to draw  the  ESSN  model (Fig.~\ref{fig:RRMSModel}b). This model is characterized by a system  of 433 ODEs, but with few assumptions it is possible to derive the corresponding reduced ODEs system including only the symbolic equations. 
In details, let us define the set of all the 27 location coordinates as $\mathbf{P}=\{ \Tuple{p_x,p_y,p_z},p_x\in\it{PosX},p_y\in\it{PosY},p_z\in\it{PosZ} \}$. Then, we consider three disjoint subsets of $\mathbf{P}$, namely $P1,\ P2,\ P3$; the first two represent the two sets grouping the EBV and DAC injection locations, respectively, and \emph{P3} the remaining locations. For simplicity and also to maintain the symmetries into the system, the  EBV and DAC injection locations do not change over the simulation time and do not overlap. Given this, it is possible to derive the symbolic ODEs (SODEs) system characterized by 49 equations instead of 433.
A further reduction is represented by the number of terms in each SODE, representative of an equivalence class of ODEs, with respect
to the number of terms appearing in the ODEs in the equivalence class, due to the factorization obtained thanks to the presence of symmetries.
Other examples are reported in Table~\ref{tab:ODEdimension}, where the R file dimension and the number of differential equations of the complete and reduced model are compared considering different cubic grid dimension, from $3\times 3\times 3$ to $5\times 5\times 5$. It is easy to see that an increasing number of locations is associated with an increase in the number of ODEs and of the R file containing them, while the SODE system does not change.  Note that when the  $5\times 5\times 5$ grid is considered, the ODEs generation procedure fails because it exceeds the available memory. \\
The advantage can also be observed from the viewpoint of the simulation time, we obtained a speed up from 8.927205 hours to 12.76043 secs on an Intel Xeon processor @ 2GHz. Note that  the simulation was performed  considering $3\times 3\times 3$ cubic grid, one year interval and assuming EBV injections at regular times (every two months), and each  injection introduces into the system 10000 EBV copies.

\begin{table}[H]
    \centering
    \begin{tabular}{c|c|c}
         Number of locations $\ $ &  R File dimension ODEs / SODEs $\ $ & Number of ODEs  / SODEs \\
         \hline
         27  (3x3x3)  & 0.43 MiB / 0.023 MiB & 433 / 49 \\
         64  (4x4x4)  & 5.0 MiB / 0.023 MiB & 1025 / 49 \\
         125 (5x5x5)  & Out of memory / 0.023 MiB &  2001 / 49 \\
         \hline
    \end{tabular}
    \caption{Comparing the ODE and SODE system, varying the cubic grid dimension.}
    \label{tab:ODEdimension}
\end{table}


\begin{figure}[ht]
\vspace{3mm}
 \begin{center}
    \includegraphics[width=1.\textwidth]{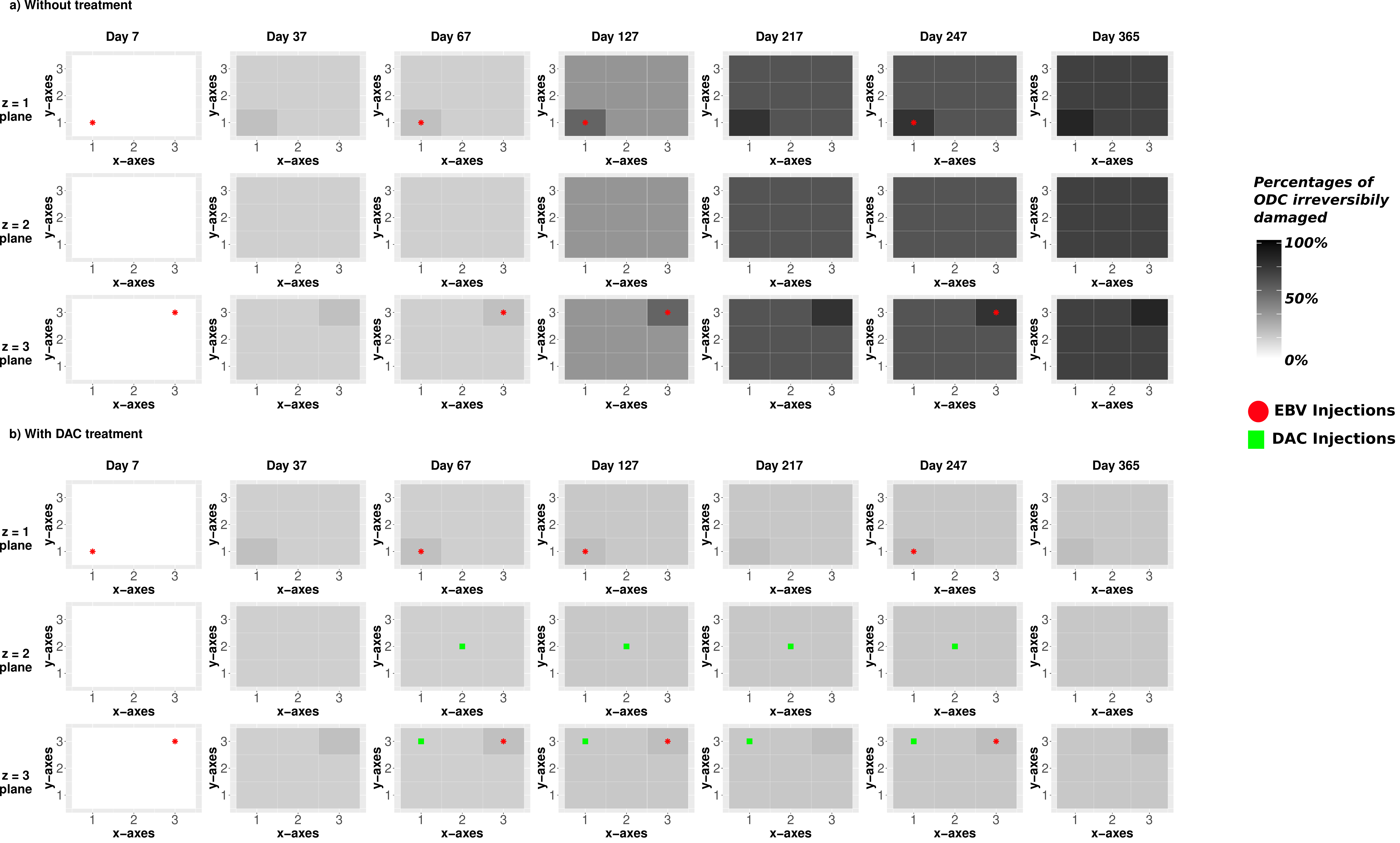}
    \caption{Percentages of ODCs irreversibly damaged a) without and b) with DAC treatment.}
    \label{fig:3Dodc}
    
 \end{center}
\vspace{-8mm}
\end{figure}

A possible evolution of the system  is shown in Fig.~\ref{fig:3Dodc}, where the red circles represent the location in which  EBV is injected.
For each plot, the three rows represent the z-planes and the columns refer to the time points in which the injections are done. Fixing the time point and the z-plane, the corresponding $3\times 3$ square reports the number of ODCs damaged into the nine grid cells obtained varying the x and y coordinates.
As expected, the panel A of Fig.~\ref{fig:3Dodc} shows the progressive accumulation of ODC irreversibly damaged until day 365.
Otherwise, in panel B of Fig.~\ref{fig:3Dodc} is reported the results of the  simulation of the DAC effect. In details, every month after two months of simulation, two injections are simulated (green squares) introducing 300 DAC copies for each administration. 
These results agree with those proposed in \cite{PerniceBMC2019} since the number of irreversibly damaged ODCs decreases in the case with DAC administration with respect to the case in which no drug is injected. With DAC the percentage of irreversibly damaged ODCs ranges from 19\% to 24\%, while with no DAC the number of irreversibly damaged ODCs is between 70\% and 85\%.




\section{\bf Conclusion}
In this work we extended the model presented in \cite{PerniceBMC2019} including the spatial coordinates of all entities in a cubic tissue portion.  This gives the opportunity to model more realistic scenarios, where different quantities of virus enter into the system from different directions.

Moreover, we described how the intrinsic symmetries of the derived ESSN model may be automatically exploited to reduce the complexity of the analysis step. This allows us to study models which are independent from the grid size, while with the classical approach it is hard to generate the ODEs system corresponding to the model with a $5\times5\times5$ grid.\\
As further work, we will focus our experiments on the dosage of DAC and on the simulation of the DAC pharmacokinetics in order to simulate the up taking of DAC by the body, its  biotransformation and the distribution of DAC and its metabolites in the tissues.



\bibliographystyle{splncs03}
\bibliography{Main.bib}

\mainmatter  

\title{Multiple Sclerosis disease: a computational approach for investigating its drug interactions.\\ \emph{Appendix} }

\titlerunning{Multiple Sclerosis disease: a computational approach for investigating its drug interactions.}

%
%

\author{Simone Pernice$^{(1)}$, Marco Beccuti$^{(1)}$, Marzio Pennisi$^{(2)}$, Giuliana Franceschinis$^{(2)}$, Gianfranco Balbo$^{(1)}$}

\institute{$\ $\\(1) Dept. of Computer Science, University of Turin, Turin, Italy\\
(2) Computer Science Inst., DiSIT, University of Eastern Piedmont, Alessandria, Italy
}

\maketitle

\section{Appendix}

In this $Appendix$ we conclude the description of the model introduced in the main paper \emph{Multiple Sclerosis disease: a computational approach for investigating its drug interactions.}, an extension of the Relapsing Remitting Multiple Sclerosis (RRMS) model presented in ~\cite{PerniceBMC2019}  considering the cells movement into a three-dimensional grid. In details, we report the function associated to the general transitions of the model and all the constants value exploited during the experiments. Let us note that in the next section we extend the functions already introduced in  ~\cite{PerniceBMC2019}, except the ones characterizing the movements, in order to consider the new color instances regarding the positions, which were not present in the previous work \cite{PerniceBMC2019}.

\subsection{General transitions}

In this section we recall that transitions which represent i) the killing of a cell, e.g., \emph{TregKillsTeff} or \emph{TeffKillsODC}, ii) the entry of cells into the system, such as \emph{EBVinj}, iii) the activation of T cells, e.g., \emph{TeffActivation}, iv) the duplication of a cell, e.g., such as \emph{TregDup}, and v) the cell movements  e.g., \emph{TregMovements}, are all modelled as general transitions because they do not follow the Mass Action (MA) law.\\
Let us recall the following notations:
\begin{itemize}
\item $f_{\Tuple{t,c}}(\hat{x}(\nu),\nu )$ is the speed of the transition $t\in T_g$ and $\hat{x}(\nu)$ represents the vector of the average numbers of tokens  for all the input places. For brevity, when the function will not depend on the color instance $c$, we will omit the $c$ and we will simplify the specification of the function in the following way  $f_{t}(\hat{x}(\nu),\nu )$.
\item $exp(-\dfrac{y}{Cost})\ $ represents a coefficient varying in $[0,1]$ and it is directly related to the $y$ value. This is usually multiplied to a constant rate, so that an increase $y$ involves a decrease in the rate associated.
\item $x_{tot_{\Tuple{p_x,p_y,p_z}}}(\nu)$ is the total number of cells  in the grid cell with coordinates $\Tuple{p_x,p_y,p_z}$ at time $\nu$. For simplicity, we will neglect all the dependencies from the position $\Tuple{p_x,p_y,p_z}$ and time $\nu$, so in the case of $x_{tot_{\Tuple{p_x,p_y,p_z}}}(\nu)$, it will be $x_{tot}$. In general the notation will be $x_{Name_{\Tuple{p_x,p_y,p_z}}}(\nu)=x_{Name}$.
\end{itemize}
All the general transitions  of the model are now explained in details and all the constants are summarized in Tables \ref{tab:constants} and \ref{tab:parameters}.
\begin{itemize}
	\item \emph{EBVinj} and \emph{DACinj} inject into the system  specific quantities of EBV and DAC respectively at fixed time points;
	
	\item \emph{FromTimoREG},  \emph{FromTimoEFF}, and \emph{NKentry} are the transitions which keep in a constant range the number of $RestingTreg$, $RestingTeff$, and $NK$ respectively. They are defined as  
			\begin{align*}
				f_{\Tuple{FromTimoREG,p_x,p_y,p_z} }(x_{ResTreg},\nu)  &= q_{RestTreg} * ( 1 -  x_{ResTreg}/63 );\\
				f_{\Tuple{FromTimoEFF,p_x,p_y,p_z}}(x_{ResTeff},\nu) &= q_{RestTeff}  * ( 1 -  x_{ResTeff}/1687 );\\
				f_{\Tuple{NKentry,p_x,p_y,p_z}}(x_{NK},\nu) &=  q_{NK}  * ( 1 -  x_{NK}/375 ),
	\end{align*}
	where $x_{ResTreg},\ x_{ResTeff},\  $ and $x_{NK}\ $ are the numbers of cells in the input places (i.e. $RestingTreg$ for \emph{FromTimoREG}, etc) at time $\nu$ and position $\Tuple{p_x,p_y,p_z}$. Then $q_{-}\ $ represents the quantity injected in the output place to preserve the cell quantity, i.e. 63 for the $RestingTreg$,  1687 for the $RestingTeff$ and 375 for the $NK$.

	\item \emph{TregActivation} and \emph{TeffActivation}  transitions model the activation of the Teff and Tref cells. In particular, these are specified as general transitions to simulate a reduced Teff activation velocity with respect to a decreasing virus presence, and a Treg activation velocity proportional to the number of Teffs and inversely proportional to the number of EBV particles (allowing the Teff to annihilate the virus).  So the functions are defined as
		\begin{align*}
 f_{\Tuple{TregActivation,p_x,p_y,p_z}}(\hat{x}(\nu),\nu) &=  r_{TregA} * \dfrac{x_{Teff} }{(x_{Teff} + x_{EBV}+ 1) }* x_{ResTreg};\\
 f_{\Tuple{TeffActivation,p_x,p_y,p_z}}(\hat{x}(\nu),\nu) &=  r_{TeffA} *  ( 1-exp(-\dfrac{x_{EBV}}{C_{EBV}})  )* x_{ResTeff},
		\end{align*}
		where $r_{TregA}$ and $r_{TeffA}$ are the activation constant rates for the Treg and Teff respectively. $\hat{x}(\nu)$, in case of the  $TregActivation$ transition, consists of the variables $x_{ResTreg},x_{EBV},x_{Teff}$ respectively to the $RestingTeff, EBV$ and $Teff$ places, differently the  $TeffActivation$ transition is characterized by $x_{ResTeff}$ and $x_{EBV} $. Finally the constant $C_{EBV}$ is related to the EBV particles and it is defined to reduce the activation rate with the decreasing of the virus presence.

	\item \emph{MemActivation} is defined as 
		\begin{equation*}
		    f_{\Tuple{MemActivation,p_x,p_y,p_z}}(\hat{x}(\nu),\nu)= \begin{cases}
		0 & \nu< t_{2inj},\\
		r_{MemA}* x_{Mem} (\nu)& \nu \ge t_{2inj},
	\end{cases}
		\end{equation*}
where
	\begin{equation*}
	    r_{MemA} =	 2 * r_{TeffA} * ( 1-\exp(- \dfrac{x_{Mem}(\nu)}{C_{Mem} }  )* ( 1- \exp(-\dfrac{x_{EBV}}{C_{EBV} } ),
	\end{equation*}
	
	 and  $t_{2inj}$ is the time corresponding to the second EBV injection. We considere this velocity as zero since the T Memory effectors start to react after the first virus occurrence. $\hat{x}(\nu)=(x_{Mem}(\nu),x_{EBV}(\nu) )$ is the marking vector storing the number of T Memory effectors (no position dependency) and EBV particles in the place $\Tuple{p_x,p_y,p_z}$, respectively at time $\nu$. While $C_{Mem}$ and $C_{EBV}$ constant related to the Memory and EBV cells needed to slow down the activation rate with the decreasing of EBV and Memory cells. This because we have to leave a minimum number of T Memory effectors into the system. So when in the system there are large number of EBV particles and of T Memory effectors, the activation speed reaches its maximum that is given by twice the velocity of the Teff cells, $r_{TeffA}$.

	\item All the transitions that model the killing of a specific cell are defined as follow:\\
	$$ \forall t \in \big\{ \emph{TregKillsTeff}, \emph{TeffKillsODC},  \emph{TeffKillsEBV}, \emph{NKKillsTcell}\big\} $$
	then
	\begin{equation*}
	f_{\Tuple{t,p_x,p_y,p_z} }(\hat{x}(\nu),\nu) = \dfrac{1}{x_{tot}} * r_{t} * \prod_{i}  \hat{x}_i(\nu) ,
	\end{equation*}
	where $\prod_{i}  \hat{x}_i(\nu) $ is the product of the average numbers of tokens in the input places of the transition $t$, $r_{t}$ is the constant rate related to the transition $t$ , and $\dfrac{1}{x_{tot}}$ represents the probability that a specific meeting between two different cells occurs in the specific grid cell.
	
	\item \emph{TregDup} transition models the Treg duplication depending proportionally on the amount of IL2 and inversely proportional on the number of DAC cells  (to simulate the reduced duplication velocity during the Daclizumab therapy), and it is defined as
		\begin{equation*}
			f_{\Tuple{TregDup,p_x,p_y,p_z}}(\hat{x}(\nu),\nu) = \eta_{TrD}(\hat{x}(\nu),\nu)* x_{Treg} * x_{IL2}* \dfrac{1}{x_{tot}},
		\end{equation*}
        with 
		\begin{equation*}
		    	\eta_{TrD}(\hat{x}(\nu),\nu ) = r_{TregDup} * ( 1-exp(-\dfrac{x_{IL2}}{C_{IL2}})  )* ( exp(-\dfrac{x_{DAC}}{C_{DAC}}) ).
		\end{equation*}
		where $r_{TregDup}$ is the constant Treg duplication rate, $\hat{x}(\nu)=\big\{x_{Treg},\ x_{IL2},\ x_{DAC} \big\}$ and  $C_{IL2}$ and $C_{DAC}$ are the constants related to the IL2 and DAC cells to slow down the duplication velocity with an increasing number of DACs and a decreasing number of IL2 proteins.
		
	\item Considering the Teff duplication event we have to distinguish two possible cases: 1) the Teff symmetric duplication with probability $p_{eff}^{dup}$  and a Teff asymmetric duplication, implying the T Memory effector differentiation,  with probability $p_{eff}^{mem}=1-p_{eff}^{dup}$ . This is modeled exploiting two different transitions: \emph{TeffDup\_Sym} and \emph{TeffDup\_Asym}. So let us define
	$$ r^{eff}_{dup} = \eta_{TeD}(\hat{x}(\nu),\nu) * x_{Teff} * x_{IL2} *\dfrac{1}{x_{tot}} $$ then these two transitions are defined as:
	\begin{align*}
   f_{\Tuple{TeffDup\_Sym},p_x,p_y,p_z}(\hat{x}(\nu),\nu) =  p_{eff}^{dup} * r^{eff}_{dup}
   \end{align*}
   and
   	\begin{align*}
   f_{\Tuple{TeffDup\_Asym,p_x,p_y,p_z}}(\hat{x}(\nu),\nu) =  p_{eff}^{mem} *r^{eff}_{dup},
	\end{align*}
	with
	\begin{equation*}
	    	 \eta_{TeD}(\hat{x}(\nu),\nu )= r_{TeffDup} * ( 1-exp(-\dfrac{x_{IL2}}{C_{IL2}})  )* ( exp(-\dfrac{x_{DAC}}{C_{DAC}}) ).
	\end{equation*}
	Where  $r_{TeffDup}$ is the constant Teff duplication rate, $\hat{x}(\nu)=\big\{x_{Teff},\ x_{IL2},\ x_{DAC} \big\}$.
\end{itemize}

For clarity,  the notation for the movements functions of the color combinations  $\Tuple{p_x,p_y,p_z}$ and $\Tuple{q_x,q_y,q_z}$, representing the location coordinates,  is simplified to $\bTuple{p}$ and $\bTuple{q}$, respectively.
\begin{itemize}
\item \textbf{\emph{TeffMovement}} simulates the movement of Teff cells from point (with coordinates represented by the color combination) $\bTuple{p}$ to point  $\bTuple{q}$. The speed of this movement (the rate of transition TeffMovement) is inversely related to the number of EBV cells in  $\bTuple{p}$ and depends on the number of EBV in  $\bTuple{q}$ such that a greater number of EBV cells leads to  a higher probability to move into that location. This is captured by the following formula
\begin{align*}
    f_{\nbTuple{TeffMovement}{p,q}} (\hat{x}(\nu),\nu)=& r_{moves} * ( exp(- \dfrac{x_{EBV_{\bTuple{p}}} }{C_{EBV}} ) ) * p^{Teff}_{\bTuple{q}} * x_{Teff_{\bTuple{p}}} 
\end{align*}
where 
$r_{moves}$ is an experimental coefficient that we set equal to $0.1$; 

$exp(- \dfrac{x_{EBV_{\bTuple{p}}} }{C_{EBV}} )$ is a term that accounts for the fact that the velocity of the movement is inversely related to the number of EBV cells in the starting point going to $0$ in a manner that is slower than $1/x_{EBV_{\bTuple{p}}}$;

$p^{Teff}_{\bTuple{q}}= \frac{x_{EBV_{\bTuple{q}}}}{EBV_{tot}}$ represents the probability to move in the cell with coordinates $\bTuple{q}$ where $EBV_{tot}$ is the total number of EBV in the grid at time $\nu$; and   $C_{EBV}$ is an experimental constant that we set equal to $1000$.
All these quantities are functions of the time $\nu$ which is omitted in the formula to keep the notation simpler.
\item \textbf{\emph{TregMovement}} represents the movements of the Treg cells from point  $\bTuple{p}$ to point $\bTuple{q}$. Similarly to what explained for transition TeffMovement the speed is inversely related to the number of Teff cells in  $\bTuple{p}$ (term  $exp(- \dfrac{x_{Teff_{\bTuple{p}}}}{C_{\it Teff} })$ ) and depends on the number of Teffs in  $\bTuple{q}$ (term  $p^{Treg}_{\bTuple{q}}= \frac{x_{Teff_{\bTuple{q}}}}{Teff_{tot}}$ ), such that a greater number of Teffs leads to  a higher probability to reach that location.

\begin{align*}
    f_{\nbTuple{TregMovement}{p,q}} (\hat{x}(\nu),\nu)=& r_{moves} * ( exp(- \dfrac{x_{Teff_{\bTuple{p}}}}{C_{\it Teff} }) ) * p^{Treg}_{\bTuple{q}} * x_{Treg_{\bTuple{p}}} 
\end{align*} 
Again, in our experiment we fixed $r_{moves}=0.1$ and $C_{\it Teff}=800$.

\item \textbf{\emph{EBVMovement}} simulates the EBV movements from point  $\bTuple{p}$ to point  $\bTuple{q}$. In this case we assume that the probability to move is equally distributed among  all the grid cells.

\begin{equation*}
    f_{\nbTuple{EBVMovement}{p,q}} (\hat{x}(\nu),\nu)= r_{moves} * p^{EBV}_{\bTuple{q}} * x_{EBV_{\bTuple{p}}} 
\end{equation*}
Also in this case, in our experiment we fixed $r_{moves}=0.1$.
\item \textbf{\emph{DACMovement}} simulates the DAC movements from point $\bTuple{p}$ to point $\bTuple{q}$. This is inversely related  to the number of T-cells (Treg+Teff) in  $\bTuple{p}$ as for the TeffMovement and TregMovement cases and directly proportional to the number of T-cells in  $\bTuple{q}$. 

\begin{align*}
    f_{\nbTuple{DACMovement}{p,q}} (\hat{x}(\nu),\nu)=& r_{moves} * ( exp(-\dfrac{x_{Tcells_{\bTuple{p}}}}{C_{Tcell}}) )* p^{DAC}_{\bTuple{q}} * x_{DAC_{\bTuple{p}}} 
\end{align*}
The quantities  $r_{moves}=0.1$ and $C_{\it Tcell}=1000$  were used in this last case.

\end{itemize}

\subsection{Parameters}

\begin{table}[H]
\centering
	\caption{List of the model fixed parameters with their corresponding values for both the scenarios: healthy and Multiple Sclerosis (MS) patient.\\}
	\label{tab:parameters}
		\begin{tabular}{c|c|c|c|}
			Transitions/events & Parameters & Healthy patient & MS patient \\ [1pt]
			\hline\hline
			\emph{Treg Death} & $ r_{TregD} $ & $1/24\ h^{-1}$ & $1/24\ h^{-1}$ \\ [3pt]
			\emph{Teff Death} & $ r_{TeffD} $ & $1/24\ h^{-1} $& $1/24\ h^{-1}$  \\ [3pt]
			\emph{NK Death}  & $ r_{NKD} $ & $ 1/24 \ h^{-1} $& $1/24\ h^{-1}$  \\ [3pt]
			\emph{NK Dup}  & $ r_{NKDup} $ & $ 1/24 \ h^{-1} $& $1/24\ h^{-1}$  \\ [3pt]
			\hline\hline 
			\emph{Teff Activation} & $r_{TeffA}$ & $0.4\ h^{-1} $ & $0.4\ h^{-1} $ \\ [3pt]
			\emph{Treg Activation} & $r_{TregA}$ & $0.2\ h^{-1} $ & $0.2\ h^{-1} $ \\ [3pt]
			\emph{Treg Dup}  & $r_{TregDup}$ & $0.09\ h^{-1} $ & $0.09\ h^{-1} $ \\ [3pt]
			\emph{Teff Dup}  & $r_{TeffDup}$  & $0.5\ h^{-1} $ & $0.5\ h^{-1} $\\ [3pt]
			\emph{TeffKillODC}  & $r_{TeKodc}$ & $0.1\ h^{-1} $ & $0.15\ h^{-1} $ \\ [3pt]
			\emph{TregKillTeff}  &$r_{TrKTe}$ & $3\ h^{-1} $ & $1\ h^{-1} $ \\ [3pt]
			\emph{TeffKillEBV}  & $r_{TeKebv}$ & $0.15\ h^{-1} $ & $0.1\ h^{-1} $ \\ [3pt]
			\emph{Recovery}  & $r_{rec}$ & $0.1\ h^{-1} $ & $0.1\ h^{-1} $ \\ [3pt]
			\emph{NKKillTcell}  &$r_{NKkTc}$ & $0.1\ h^{-1} $ & $0.1\ h^{-1} $ \\ [3pt]
			\emph{DACDeath}  & $r_{DacD}$ & $-$ & $ 1/(3*30*24) \ h^{-1} $ \\ [3pt]
			\hline
		\end{tabular}    
\end{table}

\begin{table}[H]
\centering
	\caption{List of the model constants.\\}
	\label{tab:constants}
		\begin{tabular}{|c|c|c}
			Constant &  Value \\ [1pt]
			\hline\hline
			$q_{RestTreg}$ & 20 \\ [3pt]
			$q_{RestTeff}$ & 500 \\ [3pt]
			$q_{NK}$ &  100 \\ [3pt]
			$C_{EBV}$ & 1000 \\ [3pt]
			$C_{Mem}$ & 200 \\ [3pt]
			$C_{DAC}$ & 20\\  [3pt]
			$C_{IL2}$ & 200 \\ [3pt]
			$C_{Tcell}$ & 200 \\ [3pt]
            $C_{Teff} $ & 200 \\ [3pt]
			$p_{eff}^{dup}$ &  2/3 \\ [3pt]
			$p_{eff}^{mem}$ & 1/3 \\ [3pt]
			\hline
		\end{tabular}    
\end{table}


\begin{table}[H]
\centering
	\caption{List of the cell numbers used in the model.}
	\label{tab:CellNumbers}
		\begin{tabular}{c|c|c}
			Cell & Value & Reference\\ [1pt]
			\hline\hline 
			$T Lymphocytes$ & $[3 * 10^{3} cells/mm^{3}]$ & \cite{Al-Mawali2018,warny2018} \\[3pt]
			$Resting Teff$ & $[1687 cells/ mm^{3}]$ & \cite{santagostino1999,bisset2004,choi2014,saathoff2008}\\[3pt]
			$Resting Treg$ & $[63 cells/ mm^{3}]$ & \cite{somerset2004} \\[3pt]
			$NK$ & $ [375 cells/ mm^{3}] $ & \cite{santagostino1999,bisset2004,choi2014,saathoff2008}\\[3pt]
			$ODC$ & $[125 cells/ mm^{3}]$ & \cite{segal2009} \\[3pt]
			$EBV infection$ & $[50-70 days]$ & \cite{balfour2005} \\[3pt]
			\hline
		\end{tabular}    
\end{table}


\bibliographystyle{splncs03}


\end{document}